\documentclass[fleqn,10pt,conference,a4paper]{IEEEtran}

\usepackage{amsmath,float}

\usepackage[utf8]{inputenc}
\usepackage{amsmath}
\usepackage{mathrsfs}

\usepackage[linesnumbered,ruled]{algorithm2e} 
\SetKwInOut{Input}{Input}
\SetKwInOut{Output}{Output}

\usepackage{lipsum}
\usepackage{amsmath}
\usepackage{amssymb}
\usepackage{graphicx}
\usepackage{subfig}
\usepackage[export]{adjustbox}
\usepackage{cite}
\usepackage{array}
\usepackage{ragged2e}
\usepackage{blindtext}
\usepackage{flushend}[keeplastbox] 
\usepackage{lastpage}
\usepackage{color}
\usepackage{fancyhdr}
\usepackage{url}
\usepackage{xcolor,colortbl}
\usepackage{multirow}
\usepackage{caption}
\usepackage{subcaption}
\usepackage{comment}
\usepackage[utf8]{inputenc}
\usepackage{tabularx, booktabs, makecell, caption}
\usepackage{siunitx}

\fancypagestyle{firstpage}
{
    \fancyhf{}
    \fancyhead[C]{The paper has been presented at the 2025 35th International Conference Radioelektronika (RADIOELEKTRONIKA),\\ Hnanice, Czech Republic, May 12-14, 2025, \url{https://doi.org/10.1109/RADIOELEKTRONIKA65656.2025.11008407}}
    \fancyfoot[C]{This research was funded in part by the National Science Center (NCN), Poland, grant no. 2021/43/I/ST7/03294 (MubaMilWave). For this purpose of Open Access, the author has applied a CC-BY public copyright license to any Author Accepted Manuscript (AAM) version arising from this submission.}
}

\begin{document}

\title{LSTM-based Power Delay Profile Predictions for\\ Intra-bus Wireless Propagation}
\author
{
\IEEEauthorblockN{Rajeev~Shukla\IEEEauthorrefmark{1},
Atharva~Verma\IEEEauthorrefmark{1}
Aniruddha~Chandra\IEEEauthorrefmark{1},
Ondřej Zelený\IEEEauthorrefmark{2}, 
Radek Zavorka\IEEEauthorrefmark{2},
Jiri~Blumenstein\IEEEauthorrefmark{2},\\
Ale{\v{s}}~Proke{\v{s}}\IEEEauthorrefmark{2},
Jarosław~Wojtuń\IEEEauthorrefmark{3},
Jan~M.~Kelner\IEEEauthorrefmark{3},
Cezary~Ziolkowski\IEEEauthorrefmark{3},
Domenico~Ciuonzo\IEEEauthorrefmark{4}
}
\IEEEauthorblockA{
\IEEEauthorrefmark{1}ECE Department, NIT Durgapur, 713209 Durgapur, India (aniruddha.chandra@ieee.org)
}
\IEEEauthorblockA{
\IEEEauthorrefmark{2}Department of Radio Electronics, Brno University of Technology, 61600 Brno, Czech Republic
}
\IEEEauthorblockA{
\IEEEauthorrefmark{3}Institute of Communications Systems, Faculty of Electronics, Military University of Technology, Warsaw, Poland }

\IEEEauthorblockA{
\IEEEauthorrefmark{4}Electrical Engineering and Information Technologies (DIETI), University of Naples Federico II, Naples, Campania, Italy}
}


\maketitle
\thispagestyle{firstpage}

\begin{abstract}
Long/short-term memory (LSTM) is a deep learning model that can capture long-term dependencies of wireless channel models and is highly adaptable to short-term changes in a wireless environment. This paper proposes a simple LSTM model to predict the channel transfer function (CTF) for a given transmitter-receiver location inside a bus for the $60$ GHz millimetre wave band. The average error of the derived power delay profile (PDP) taps, obtained from the predicted CTFs, was less than $10\%$ compared to the ground truth. 
\end {abstract}



\section{Introduction}\label{sec:intro}
Channel characterisation is critical to the design and optimisation of wireless communication systems. It helps to understand the propagation of signals across the wireless channel and mitigate the effect of fading, interference, delay, and pathloss. Channel characterisation using real-time measurement of channel characteristics (like pathloss, fading, power delay profile, delay spread, etc.) prompts the development of empirical, deterministic, and artificial intelligence (AI)-based channel models. These models are beneficial in the event of unavailability of measuring equipment, limited data sources, and lack of financial means, which makes generalising different channel environments difficult \cite{Yin_CC_2016}. Empirical channel models are not dynamic, often neglect site-specific effects, and have limited accuracy in complex environments. AI and machine learning (ML), especially deep learning (DL)--based channel models, offer promising solutions for dynamic and complex channel environments. DL models provide more accurate predictions by learning from complex propagation patterns from real-time data, thereby adapting to changes in complex physical environments \cite{Seretis_NN_2022, Seretis_ML_2022}. 

\section{Proposed Methodology}

\begin{figure*}[hbt!]
    \centering
    \includegraphics[scale=0.13]{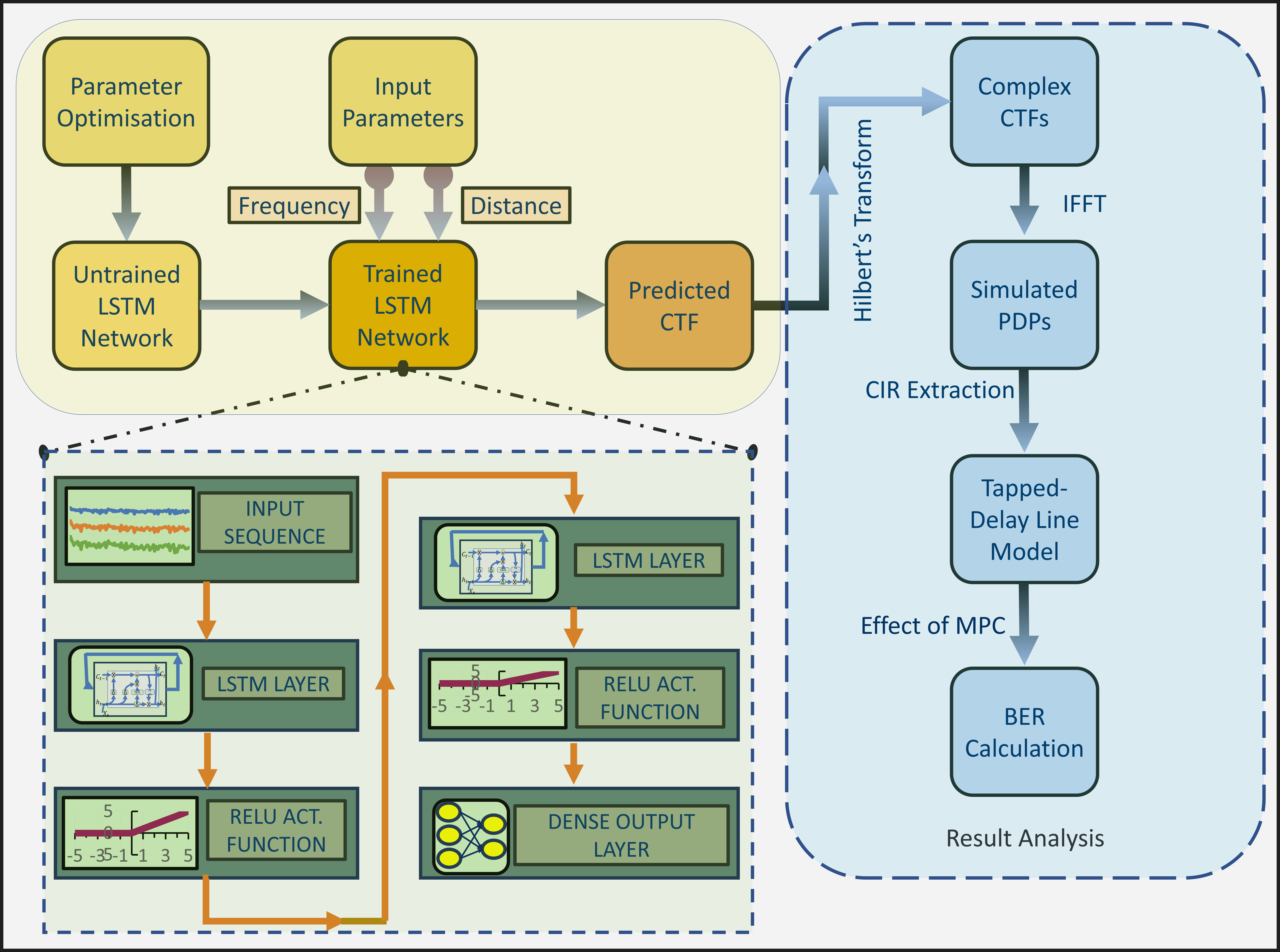}
    \caption{Flow chart for CTF generation using a simple 2-layer LSTM model.}
    \label{Fig.1}
\end{figure*}

With the advent of more connected and autonomous vehicles, channel characterisation inside a vehicle plays a pivotal role in optimising wireless channel performance. Highly reflective metal surfaces and a cluttered physical environment cause significant multipath effects impacting signal strength and delay spread. There had been a renewed interest in millimetre-wave (mmWave) communication with unconventional use cases emerging recently \cite{sambhwani2025extending}. The $60$GHz mmWave band made popular with IEEE 802.11ad/ay standards, allows multi-gigabyte wireless connectivity and uses compact high-gain antennas for wireless connectivity \cite{Website_60GHz_2025}. However, the $60$GHz band has limited penetration capability, which prevents external interference and thus provides seamless connectivity inside public transport vehicles like buses and trains or autonomous vehicles. Due to frequent signal variations and pronounced multipath components (MPCs), traditional analytical methods fail to capture the complex dynamic channel. Thus, AI-driven models can analyse real-world measurement data to predict PDPs or CTFs and improve the accuracy of tapped delay line (TDL) models or ray-tracing predictions.  

AI/ML can revolutionise channel characterisation and channel modelling for the modern wireless communication system. The authors in paper \cite{Huang_AI_CC_2022_1, Huang_AI_CC_2022_2} argued in support of this claim. The authors first provided a comprehensive survey on using ML in channel characterisation and antenna optimisation in these papers. Then, they detailed channel scenario identification and channel modelling. Similar assertions were made in \cite{Pham_AI_2021}, where the authors explored the integration of AI and visible light communication to provide assistive solutions to improve navigation, avoid obstacles, and improve the overall quality of life of visually impaired people. The paper \cite{Kai_ML_CE_2021} gave an in-depth analysis of ML-based channel estimation for orthogonal frequency division multiplexing (OFDM) systems for limited training data. The findings in \cite{Shehzad_DL_CP_2022} suggested a recurrent neural network (RNN) hybrid with NeuralProphet to enhance the efficiency and reliability of forecasting channel state information. To make the AI applications more transparent and trustworthy, authors in \cite{Gizzini_XAI_CE_2024} proposed the integration of explainable AI (XAI) with DL models. The authors suggested an XAI-based channel estimation (XAI-CHEST) to estimate doubly-selective channels.\par

Advancements in intelligent transportation systems, increasing autonomy of vehicles, and high bandwidth requiring user-friendly experience make it crucial to investigate in-vehicle wireless channel characteristics. However, studies on in-vehicle channel characterisation are limited to only a few specific scenarios. The most prominent direction of studies is private vehicles like cars and vans and public vehicles like buses and train wagons. In \cite{Matolak_InV_2012}, authors presented $5$GHz band channel characterisation inside a bus and a mini-van, whereas \cite{Azpilicueta_bus_2015} studied the effect of topological impact on wireless sensor networks (WSNs) performance. The authors used an in-house developed 3D ray-tracing tool to study the channel behaviour in the absence and presence of human beings inside the bus for proper optimisation of the transceiver position. In another such use-case scenario, extensive real-time measurement \cite{chandraBus}, the authors measured the channel transfer function (CTF) inside a bus using $55$GHz - $65$GHz millimetre waves. Notably, these studies are confined to empirical and deterministic channel characterisation techniques and cannot thus adapt to any physical changes and are non-generalisable.


\section{LSTM-based CTF from Measurement Data}

The current authors investigated the behaviour of wireless channels inside a bus in an extensive measurement campaign at Brno University, Czech Republic, and reported it in \cite{chandraBus, ACBusAnalysis_2019, RahmanBus_2016}. Analysis of measurements suggested that traditional methods could not accurately capture the highly clustered intra-vehicular channel environment and thus became less effective in interpolating and extrapolating \cite{Shukla_ANN_2023} PDP beyond the measured data set.\par

LSTMs are designed for sequential data and can capture temporal dependencies. CTFs for mmWaves are highly frequency-dependent due to their highly dispersive nature. LSTMs can learn the correlation between CTFs and frequency and capture variations across different transmitter-receiver ($T_X$-$R_X$) distances in the intra-vehicular cluttered channel scenario. Motivated by the need for accurate channel modelling at millimetre-wave (mmWave) frequencies, the authors propose an LSTM-based CTF prediction model to forecast CTFs in the 55–65 GHz mmWave band. This LSTM-based prediction follows a supervised learning approach, utilising measurement data from the authors' own measurement campaigns reported in [14]–[16]. The dataset comprises CTF measurements collected over various $T_X$-$R_X$ distances within the 55–65 GHz frequency range. The predicted CTFs are analysed using the TDL model, which is subsequently employed for bit-error-rate (BER) calculation.
\begin{algorithm}[!ht]
\DontPrintSemicolon
\For{$layer1 \gets 20$ to $100$}
{
 \For{$layer2 \gets 2$ to $9$} 
  {
   \textbf{Initialization}: Random $seed$, $model \gets None$, epochs $\gets e$\;
            \textbf{Create} $layer1$: Add LSTM layer, activate ReLU, enable return sequence\;
            \textbf{Create} $layer2$: Add LSTM layer, activate ReLU, disable return sequence\;
            \textbf{Create dense layer}: Add output dense layer with 1 unit\;
            \textbf{Compile} $model$:  MSE loss, Adam optimiser\; 
            \textbf{Set}: $batch\_size\gets 20$, disable $shuffle$\;
            \textbf{Train} $model$\;
            \textbf{Calculate}: $loss\_train$, $loss\_test$\;
 }
}
\textbf{Select}: \{$layer1$, $layer2$, $e$\}$|_{min: loss\_train, loss\_test}$\;
\caption{Hyperparameter tuning for LSTM model}\label{alg:lstm_tuning}
\end{algorithm}

The LSTM model requires sequential data for prediction. The proposed LSTM model takes a simplistic approach to predicting the CTFs using only two physical parameters as input. The first parameter for the LSTM model is the frequency parameter, where $55$ GHz-$65$ GHz frequency band with a step size of $10$ MHz is taken to learn the sequential variations in the data, as we have a data point for each frequency step. The measurements were taken for various $T_X$-$R_X$ distances, $d \in$ \{$1.18$m, $1.47$m, $1.66$m, $2.24$m, $2.35$m, $3.66$m, $5.12$m, $5.16$m, $6.66$m, $6.76$m, $8.18$m, $9.36$m and $9.72$m\}, to capture the variations across distances (refer to top-right corner of Fig.\ref{Fig.1}). 

\begin{figure*}[] 
    \centering
    \subfloat[\centering $T_X$-$R_X$ distance = 3.7m.]{\includegraphics[scale=0.33]{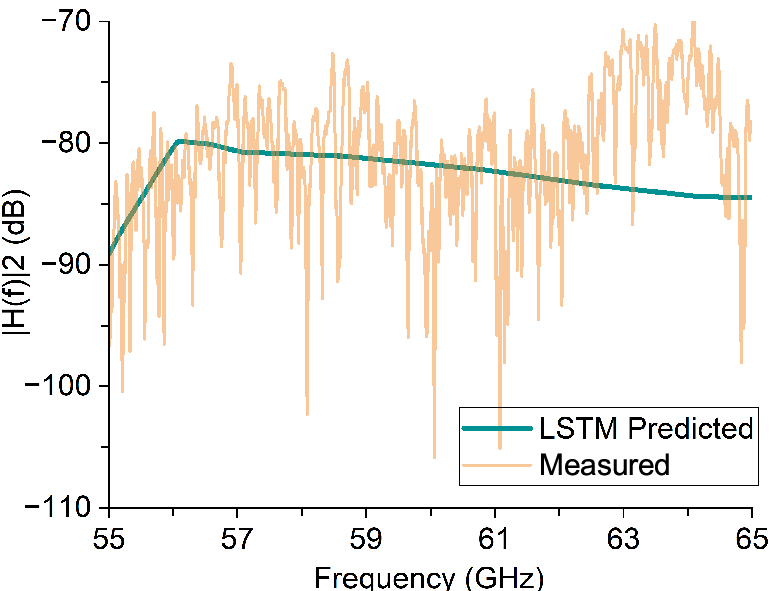}}
     \subfloat[\centering $T_X$-$R_X$ distance = 9.75m.]{\includegraphics[scale=0.33]{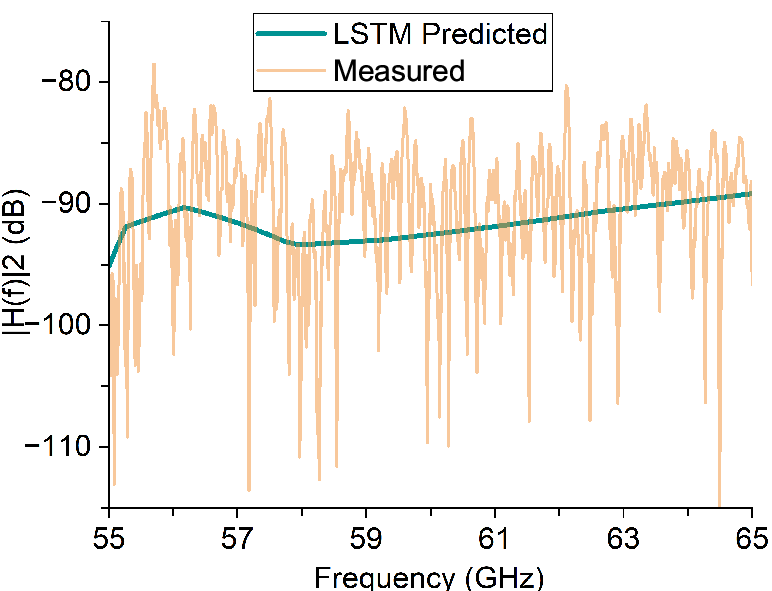}}
    \caption{Comparison between measured channel gain and LSTM predicted CTFs at different $T_X$-$R_X$ distances.}
    \label{Fig.2}%
\end{figure*}

\begin{figure*} []
    \centering
    \subfloat[\centering $T_X$-$R_X$ distance = 3.7m.]{\includegraphics[scale=0.33]{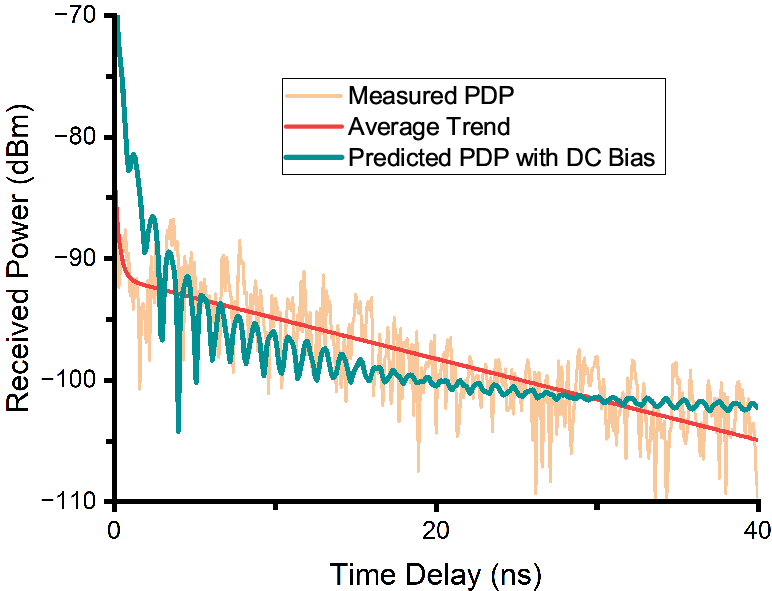}}
     \subfloat[\centering $T_X$-$R_X$ distance = 9.75m.]{\includegraphics[scale=0.33]{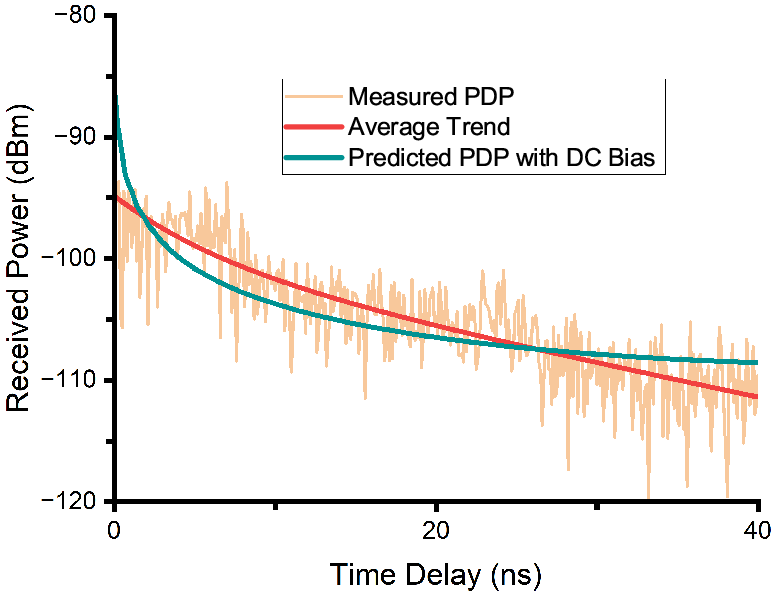}}
    \caption{Comparison between measured and LSTM predicted PDPs at different $T_X$-$R_X$ distances.}
    \label{Fig.3}%
\end{figure*}

\begin{figure*} []
    \centering
    \subfloat[\centering $T_X$-$R_X$ distance = 3.7m.]{\includegraphics[scale=0.22]{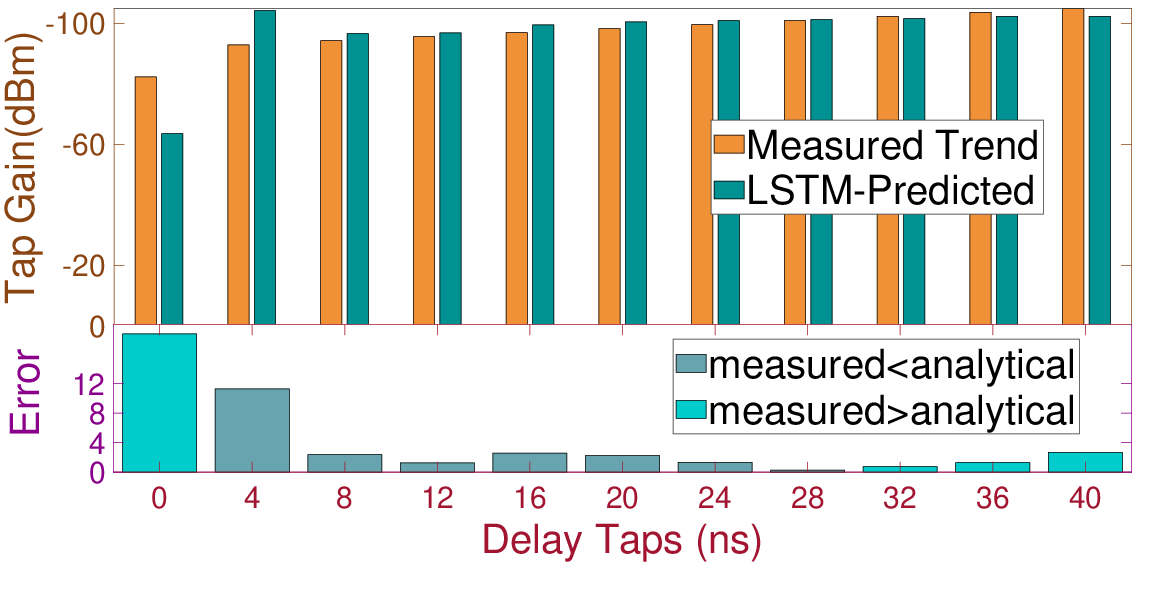}}
     \subfloat[\centering $T_X$-$R_X$ distance = 9.75m.]{\includegraphics[scale=0.22]{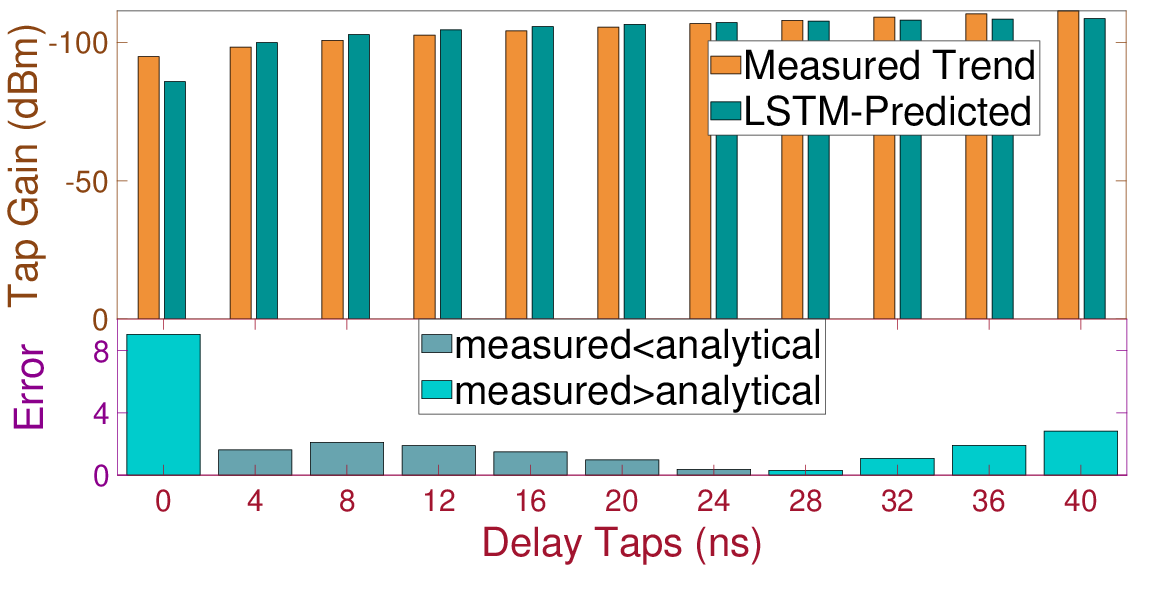}}
    \caption{Comparison between measured and LSTM predicted TDLs at different $T_X$-$R_X$ distances.}
    \label{Fig.4}%
\end{figure*}

\begin{figure*} []
    \centering
    \subfloat[\centering $T_X$-$R_X$ distance = 3.7m.]{\includegraphics[scale=0.33,trim= 0cm 0cm 0cm 0cm]{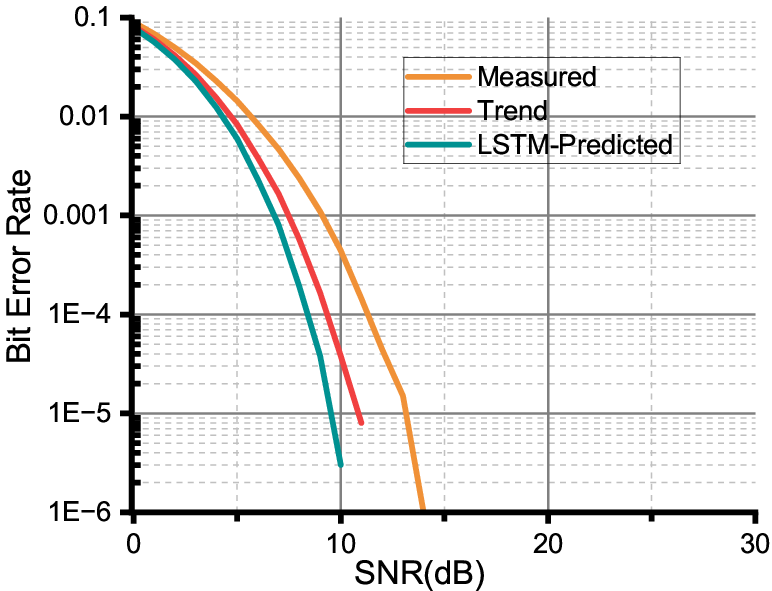}}
     \subfloat[\centering $T_X$-$R_X$ distance = 9.75m.]{\includegraphics[scale=0.33,trim= 0cm 0cm 0cm 0cm]{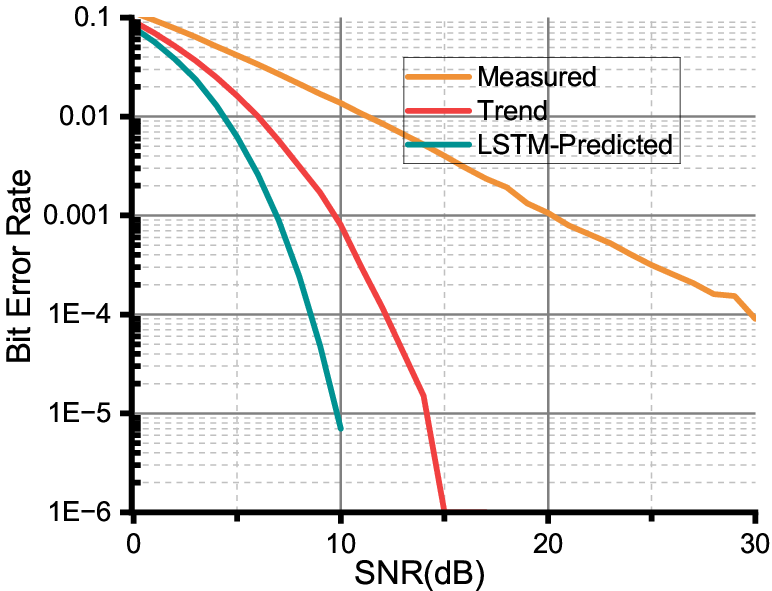}}
    \caption{Comparison between measured and LSTM predicted BER at different $T_X$-$R_X$ distances.}
    \label{Fig.5}%
\end{figure*}

Next, we describe the methodology of hyperparameter tuning through Algorithm \ref{alg:lstm_tuning}. We opted for two LSTM layers, where the first LSTM returns sequences, and the second layer is to learn from the sequence returned by the first layer and returns a single output per cell to the third ANN layer. After optimisation, the combination of hyperparameters with minimum test loss was considered. 

As far as the architectural specifications are concerned, the number of LSTM cells in the first and second layers were 100 and 9, respectively. On the other hand, the training specifications are as follows. We used Adam optimizer, and the number of epochs and batch size were decided to be 94 and 20. 

After proper hyperparameter tuning, the tuned LSTM network is trained using the training datasets and the trained network is tested on two unused databases. The test subjects have distances of $3.7$m and $9.75$m, implicating the usefulness of the LSTM-based model for interpolation and extrapolation for new frequency-distance pairs.


\section{Results and Discussion}
The authors used the proposed LSTM model to predict the CTFs from variations in frequency and distance. Figs. \ref{Fig.3}, \ref{Fig.4}, and \ref{Fig.5}, illustrate a comparative analysis between measured and predicted CTFs, PDPs and TDL models for different $T_X$-$R_X$ distances.

The comparison between measured CTFs and predicted CTFs in Fig. \ref{Fig.2} shows the ability of LSTM to follow the general trend of measured CTF data, especially in the lower frequency regions. The measured data's general trend represents the multipath components' general decay behaviour over time delay and is helpful for modelling and performance evaluation purposes. The trend captures the overall shape of the PDP, suppressing the influence of small-scale fading and measurement noise. Notably, the predicted CTFs smoothen the fluctuations in measured CTFs caused by the multipath effect of the channel environment. The CTF values from measured and predicted CTF data generate channel impulse response (CIR), as explained in Fig. \ref{Fig.1}. The PDP values from the LSTM model preserve the power distribution trend of the measured data in Fig. \ref{Fig.3}. Similar to the case of CTFs, LSTM-based PDPs try to smooth the high-frequency multipath fading effects. The predicted PDP curve retains the channel characteristics, though some deviations can be observed at higher delays due to multipath fading and noise. Fig. \ref{Fig.4} further analyses the CIRs for the LSTM model in terms of the TDL model. The bottom error plot between the CIRs of the measured average trend and the LSTM-predicted shows an acceptable deviation $(\leq 10 \%)$ at different delay taps, validating the ability of LSTM to capture the CIRs. The error matrix in table \ref{table:1} supports the validation.

The TDL model is applied to simulate the channel, and BER is calculated in Fig. \ref{Fig.5}. The graphs show that the BER drop rate gets slower for the measured TDL as the distance between $T_X$-$R_X$ increases. This is probably due to stronger multipath and deep fades, which results in a more extensive delay spread. The BER from the LSTM model is closer to the measured TDL for smaller distances. This indicates that a more substantial line-of-sight component and lower delay spread contribute to overcoming multipath effects. 

The BER analysis and error matrix signifies that LSTM successfully learned the channel characteristics and can be used for new $T_X$-$R_X$ pairs. The results also show that the performance of the presented model weakens in severe fading conditions, and more training datasets and training features can be included for better learning. 
\begin{table}[hbt!]
\caption{\small Average error across the taps of the PDPs (measured vs. trend, measured vs. predicted, and trend vs. predicted).}
\begin{center}
    \begin{tabular}{|c|c|c|c|c|}
    \hline
    \multirow{2}{*}{Type of PDPs compared} & \multicolumn{2}{c|}{$d = 3.7$ m} & \multicolumn{2}{c|}{$d = 9.75$ m} \\\cline{2-5}
    
    & RMSE & $R^2$ & RMSE & $R^2$  \\
    \hline
    \emph{measured} vs. \emph{trend} & $3.12$ & $0.84$ & $1.88$ & $0.95$ \\
     \hline
    \emph{measured} vs. \emph{predicted} & $7.96$ & $0.47$ & $3.04$ & $0.78$ \\
     \hline
    \emph{trend} vs. \emph{predicted} & $6.80$ & $0.68$ & $3.14$ & $0.77$ \\
    \hline
\end{tabular}
\end{center}
\label{table:1}
\end{table}%

\section{Conclusion}

A simple LSTM-based model is presented here to predict CTFs in an intra-vehicular channel environment. The proposed model learned from highly clustered complex channel conditions inside the bus and could predict CTFs for unmeasured $T_X$-$R_X$ distances. PDP, TDL and BER-based analysis between the measured and predicted CTFs are provided to support the proposed model. The authors observed that the results are close to the average trend of the measured data, indicating that such models are suitable for interpolation and extrapolating new $T_X$-$R_X$ distance points.\par 

\section*{Acknowledgement}
This work was developed within a framework of the research grants: project no. 23-04304L sponsored by the Czech Science Foundation, MubaMilWave no. 2021/43/I/ST7/03294 funded by the National Science Centre, Poland, under the OPUS call in the Weave programme, grant no. UGB/22-748/2024/WAT sponsored by the Military University of Technology, and chips-to-startup (C2S) program no. EE-9/2/2021-R\&D-E sponsored by MeitY, Government of India.

\bibliographystyle{./IEEEtran}    
\bibliography{IEEEabrv,./References}

@string{vtc = "Proc.\ {VTC}"}

@string{globecom = "Proc.\ Globecom"}

@INBOOK{Yin_CC_2016,
  author={Yin, Xuefeng and Cheng, Xiang},
  booktitle={Propagation Channel Characterization, Parameter Estimation, and Modeling for Wireless Communications}, 
  title={Characterization of Propagation Channels}, 
  year={2016},
  publisher = {Wiley},
  volume={},
  number={},
  pages={15-40},
  }

@ARTICLE{Seretis_NN_2022,
  author={Seretis, Aristeidis and Sarris, Costas D.},
  journal={IEEE Trans. Antennas Propag}, 
  title={Toward Physics-Based Generalizable Convolutional Neural Network Models for Indoor Propagation}, 
  month     = "Jun.",
  year={2022},
  volume={70},
  number={6},
  pages={4112-4126},
  }

@ARTICLE{Seretis_ML_2022,
  author={Seretis, Aristeidis and Sarris, Costas D.},
  journal={IEEE Trans. Antennas Propag}, 
  title={An Overview of Machine Learning Techniques for Radiowave Propagation Modeling}, 
  month     = "Jun.",
  year={2022},
  volume={70},
  number={6},
  pages={3970-3985},
  }

@electronic{Website_60GHz_2025, 
  author       = "{Research and Markets}",
  title        = "Global {WiGig} Industry booms: Next-gen wireless connectivity fuels 27.5\% CAGR through 2029",
  month        = "Jan.",
  year         = "2025",
  howpublished = "{GlobeNewswire}", 
  URL          = {https://www.globenewswire.com/news-release/2025/01/03/3003840/28124/en/Global-WiGig-Industry-Booms-Next-Gen-Wireless-Connectivity-Fuels-27-5-CAGR-Through-2029.html}
}

@article{sambhwani2025extending,
  title     = "Extending {mmWave} Deployment in the Next-Generation Network{: Coverage} and Reliability Enhancements",
  author    = "Sambhwani, Sharad and Chae, Hyukjin and Chiu, Sung-En and Fan, Boqiang and Sarkas, Ioannis and Sun, Wanlu and Zhou, John and Zhu, Xipeng",
  journal   = "IEEE Wireless Commun.",
  volume    = "32",
  number    = "1",
  pages     = "83--89",
  month     = "Feb.",
  year      = "2025"
}

@inproceedings{chandraBus,
  title     = "{60 GHz mmW channel measurements inside a bus}",
  author    = "Chandra, Aniruddha and Mikulasek, Tomas and Blumenstein, Jiri and Prokes, Ales",
  booktitle = "{Proc. IEEE IFIP NTMS}",
  address   = "Larnaca, Cyprus",
  pages     = "1--5",
  month     = "Nov.",
  year      = "2016"
}

@ARTICLE{Huang_AI_CC_2022_1,
  author={Huang, Chen and He, Ruisi and Ai, Bo and Molisch, Andreas F. and Lau, Buon Kiong and Haneda, Katsuyuki and Liu, Bo and Wang, Cheng-Xiang and Yang, Mi and Oestges, Claude and Zhong, Zhangdui},
  journal={IEEE Transactions on Antennas and Propagation}, 
  title={Artificial Intelligence Enabled Radio Propagation for Communications—Part I: Channel Characterization and Antenna-Channel Optimization}, 
  year={2022},
  volume={70},
  number={6},
  pages={3939-3954},
  }

@ARTICLE{Huang_AI_CC_2022_2,
  author={Huang, Chen and He, Ruisi and Ai, Bo and Molisch, Andreas F. and Lau, Buon Kiong and Haneda, Katsuyuki and Liu, Bo and Wang, Cheng-Xiang and Yang, Mi and Oestges, Claude and Zhong, Zhangdui},
  journal={IEEE Transactions on Antennas and Propagation}, 
  title={Artificial Intelligence Enabled Radio Propagation for Communications—Part II: Scenario Identification and Channel Modeling}, 
  year={2022},
  volume={70},
  number={6},
  pages={3955-3969},
  }

@ARTICLE{Pham_AI_2021,
  author={Pham, Quoc-Viet and Nguyen, Nhan Thanh and Huynh-The, Thien and Bao Le, Long and Lee, Kyungchun and Hwang, Won-Joo},
  journal={IEEE Access}, 
  title={Intelligent Radio Signal Processing: A Survey}, 
  year={2021},
  volume={9},
  number={},
  pages={83818-83850},
  }

@ARTICLE{Kai_ML_CE_2021,
  author={Mei, Kai and Liu, Jun and Zhang, Xiaochen and Rajatheva, Nandana and Wei, Jibo},
  journal={IEEE Transactions on Communications}, 
  title={Performance Analysis on Machine Learning-Based Channel Estimation}, 
  year={2021},
  volume={69},
  number={8},
  pages={5183-5193},
  doi={10.1109/TCOMM.2021.3083597}}

@INPROCEEDINGS{Shehzad_DL_CP_2022,
  author={Shehzad, M. Karam and Rose, Luca and Azam, M. Furqan and Assaad, Mohamad},
  booktitle={GLOBECOM 2022 - 2022 IEEE Global Communications Conference}, 
  title={Real-Time Massive MIMO Channel Prediction: A Combination of Deep Learning and NeuralProphet}, 
  year={2022},
  volume={},
  number={},
  pages={1423-1428},
  }

@ARTICLE{Gizzini_XAI_CE_2024,
  author={Gizzini, Abdul Karim and Medjahdi, Yahia and Ghandour, Ali J. and Clavier, Laurent},
  journal={IEEE Transactions on Vehicular Technology}, 
  title={Towards Explainable AI for Channel Estimation in Wireless Communications}, 
  year={2024},
  volume={73},
  number={5},
  pages={7389-7394},
  }

@INPROCEEDINGS{Matolak_InV_2012,
  author={Matolak, David W. and Chandrasekaran, Arvind},
  booktitle={2012 IEEE Vehicular Technology Conference (VTC Fall)}, 
  title={5 GHz Intra-Vehicle Channel Characterization}, 
  year={2012},
  volume={},
  number={},
  pages={1-5},
  address   = {Larnaca, Cyprus},
  }

@ARTICLE{Azpilicueta_bus_2015,
  author={Azpilicueta, Leire and López Iturri, Peio and Aguirre, Erik and Astrain, José Javier and Villadangos, Jesús and Zubiri, Cristobal and Falcone, Francisco},
  journal={IEEE Transactions on Intelligent Transportation Systems}, 
  title={Characterization of Wireless Channel Impact on Wireless Sensor Network Performance in Public Transportation Buses}, 
  year={2015},
  volume={16},
  number={6},
  pages={3280-3293},
  }

@ARTICLE{ACBusAnalysis_2019,  
author={Chandra, Aniruddha and Rahman, Aniq Ur and Ghosh, Ushasi and García-Naya, José A. and Prokeš, Aleš and Blumenstein, Jiri and Mecklenbräuker, Christoph F.}, journal={IEEE Access},   
title={60-GHz Millimeter-Wave Propagation Inside Bus: Measurement, Modeling, Simulation, and Performance Analysis},   
year={2019},  
volume={7},  
number={},  
pages={97815-97826},  
}

@inproceedings{RahmanBus_2016,
  title     = "{Channel Modelling for 60GHz mmWave Communication Inside Bus}",
  author    = "Rahman, Ainq Ur and Ghosh, Ushasi and  Chandra, Aniruddha and Prokes, Ales",
  booktitle = "{Proc. 2018 IEEE VNC}",
  address   = "Taipei, Taiwan",
  pages     = "1--6",
  month     = "Dec.",
  year      = "2018"
}

@ARTICLE{Shukla_ANN_2023,
  title    = "A simple {ANN-MLP} model for estimating               {60-GHz} {PDP} inside
              public and private vehicles",
  author   = "Shukla, Rajeev and Sarkar, Abhishek                    Narayan and Chandra, Aniruddha and Kelner,             Jan M and Ziolkowski, Cezary and Mikulasek,
              Tomas and Prokes, Ales",
  journal  = "EURASIP Journal on Wireless Communications and Networking",
  volume   =  2023,
  number   =  1,
  pages    = "47",
  month    =  Jun,
  year     =  2023
}


\end{document}